\documentclass[10pt]{iopart}
\usepackage{amsfonts}
\usepackage{amssymb}
\def\a{\alpha} 
\def\b{\beta}
\def\g{\gamma}

 \eqnobysec
 \begin{document}

 \title{On light propagation in premetric electrodynamics. 
Covariant dispersion relation}

\author{Yakov Itin}

\address{Institute of Mathematics, Hebrew University of
   Jerusalem \\
   and Jerusalem College of Technology}
\ead{itin@math.huji.ac.il}
\begin{abstract}
The premetric approach to electrodynamics provides a unified description of a wide class of electromagnetic phenomena. In particular, it involves  axion, dilaton and skewon modifications of the classical electrodynamics. This formalism emerges also when the non-minimal coupling between the electromagnetic tensor and  the torsion of Einstein-Cartan gravity is considered. Moreover, the premetric formalism can serve as a general covariant background of the electromagnetic properties of  anisotropic media. 
In the current paper, we study wave propagation in the premetric electrodynamics. 
We derive a system of characteristic equations corresponded to premetric generalization of the Maxwell equation.  
This singular system is characterized by the adjoint matrix which turns to be of a very special form --- proportional to a scalar quartic factor. 
We prove that a necessary condition for existance  a non-trivial solution of the characteristic system is expressed by a unique  scalar dispersion relation.  In the tangential (momentum) space, it determines a fourth order light hypersurface
which replaces the ordinary light cone of the standard Maxwell theory. We derive an explicit form of
the covariant dispersion relation and establish it's
algebraic and physical  origin.
\end{abstract}
\section{Introduction}
On classical and quantum levels, Maxwell's
electrodynamics is a well established theory, which results are in
a very precise coordination with the experiment. This theory,
however, can require some principle modifications in order to
include non-trivial interactions with other physical fields. The
following  non-complete list  indicates some 
directions of the possible alternations:
\begin{itemize}
\item {\it Dilaton field.} This scalar partner of the classical
 electromagnetic field is proposed recently as a source of a possible
  variation of the fine-structure constant \cite{Bekenstein:1982eu},
  \cite{Bekenstein:2002wz}.

\item{\it Axion field.} This pseudo-scalar field is believed to play a
 central role in violation of Lorentz and parity invariance
 \cite{Carroll:1989vb} --- \cite{Itin:2007cv}.

\item{\it Birefringence and optical activity of vacuum.} These effects are 
forbidden in the standard (minimal coupling) model of interaction between the
electromagnetic and the gravitational fields. In the Cartan-Einstein
model of gravity, the non-minimal coupling yields, in general, the
nontrivial effects of electromagnetic wave propagation
\cite{Preuss:2004pp} --- \cite{Preuss:2005ic}.
\end{itemize}
Although, the mentioned problems belong to rather  different branches
 of the classical field theory, their joint treatment can be
 provided in a unique framework of the {\it premetric electrodynamics}.
The roots of such an approach can be found in  the older
literature \cite{Post62}, but its final form was derived
 only recently, see \cite{ROH2002}-\cite{HIO} and specially the
 book \cite{Birkbook} and the references given therein.

In the premetric  construction, the electromagnetic field is
considered on a bare differential manifold without metric and/or
connection. Instead, the manifold is assumed to be endowed with a
fourth order constitutive pseudo-tensor, which provides the
constitutive relation between the electromagnetic field strength
and  the electromagnetic excitation tensors. The metric tensor
itself is only a secondary quantity in this construction. Its explicit form
and even its  signature is derived from the properties of the
constitutive tensor,  \cite{Obukhov:2000nw} --- \cite{Itin:2004qr}.

In the current paper, we study the electromagnetic waves
propagation in the premetric electrodynamics. From the technical
point of view, our approach is similar to those used in the
relativistic plasmodynamics \cite{Melrose1} --- \cite{Melrose3}. 
A principle different that we are dealing with a metric-free background, thus the norm
of the wave covector and its scalar product with another covector
are not acceptable. Roughly speaking,  the indices in the
tensorial expressions cannot be raised or downed. Although this restriction complicates  some derivations, it is clearly necessary in the framework of the premetric approach. Moreover, it shows that, for the electromagnetic waves propagation, the metric tensor is indeed a secondary structure. The metric structure must be considered as a result of the properties of the wave propagation and not as a predeclared fact. 

The main result of our consideration is a rigorous derivation of the covariant dispersion relation. It is shown to be originated in the adjoint matrix of a characteristic system of the generalized Maxwell field equations.

 The organization of the paper is as follows: In Section 1, we give a
brief account of the premetric electrodynamics formalism. Section
2 is devoted to the geometric optics approximation and the wave-type
ansatz. When this ansatz is substituted in the Maxwell
system, the former is transformed into a algebraic system of
linear equations. The algebraic features of this characteristic system is
studied in Section 4. A covariant dispersion relation emerges in
Section 5 as a necessary condition for existence a non-trivial
wave-type solutions in premetric Maxwell electrodynamics.  In Section 6, we give an application of the developed
formalism to the simplest Maxwell case. 
 The standard expressions of the Maxwell theory are 
reinstated. Section 7 is devoted to a discussion of the proposed
formalism and its possible generalizations.

\section{Premetric electrodynamics formalism}
 \subsection{Motivations}
In order to represent the  motivations of the premetric
electrodynamics, we briefly recall some electromagnetism models
which naturally lead to this generalization.

\vspace{0.5cm}

\noindent{\it A. Vacuum electrodynamics.} In the flat Minkowski
spacetime, electromagnetic field  is described by an antisymmetric
tensor of the electromagnetic field strength $F_{ij}$.
 In an orthogonal Cartesian coordinate system $\{x^i\}$ with $i=0,1,2,3$,
  the dynamics of the field is defined from a  pair of the first order
  partial differential equations
   \begin{equation}\label{Max0}
\epsilon^{ijkl}F_{jk,l}=0\,,\qquad F^{ij}{}_{,j}=J^i\,.
  \end{equation}
Here, the comma denotes the partial differentiation. The Levi-Civita
permutation pseudotensor $\epsilon^{ijkl}$ with  the values $\{-1,0,1\}$
is normalized by $\epsilon^{0123}=1$.

The first equation of  (\ref{Max0}) is completely independent of
the metric. In the second one, the  Minkowski metric,
$\eta^{ij}={\rm diag}(-1,1,1,1)$, is involved implicitly. It is
used here  for definition of the covariant components of the field
strength, i.e, for raising the indices
    \begin{equation}\label{Max01}
F^{ij}=\eta^{im}\eta^{jn}F_{mn}\,.
 \end{equation}
 To have a representation similar to those used below, we rewrite this equation as 
   \begin{equation}\label{Max01x}
F^{ij}=\frac 12\chi^{ijmn}F_{mn}\,,\quad {\rm  where}\quad 
\chi^{ijmn}=\eta^{im}\eta^{jn}-\eta^{in}\eta^{jm}\,.
   \end{equation}
  In (\ref{Max0}), the vector field $J^i$ describes the electric current.
  Since the tensor $F^{ij}$ is antisymmetric, the electric charge conservation law
\begin{equation}\label{Max02}
J^i{}_{,i}=0\,
   \end{equation}
is a straightforward consequence of (\ref{Max0}).

The relations above  are invariant under a 
subgroup of linear rigid transformations of coordinates which
preserves the specific form of the Minkowski metric. This group
includes the instantaneous spatial rotations, Lorentz's
transformations and reflections.

\vspace{0.5cm}

 \noindent{\it B. Electrodynamics in gravity field.}
In  a non-inertial frame, i.e., in curvilinear coordinates on the flat  spacetime, the  Minkowski metric
 $\eta^{ij}$ is  replaced by a generic pseudo-Riemannian metric
  $g^{ij}$ which components depend on a spacetime point.
On this background, the transformational requirements are changed. 
The field equations  must be now invariant
under arbitrary smooth transformations of the coordinates. To
satisfy this transformational requirement, the field equations
(\ref{Max0}) are modified to
   \begin{equation}\label{Max1}
\epsilon^{ijkl}F_{jk,l}=0\,,\qquad \left(F^{ij}\sqrt{-g}\right)_{,j}=
  J^i\sqrt{-g}\,,
  \end{equation}
  where $g={\rm det}\left(g_{ij}\right)$. The covariant components
  of the electromagnetic field strength are defined now via a multiplication
  by the metric tensor components which, in contrast to (\ref{Max01}),
   depend on a point
    \begin{equation}\label{Max1-1}
F^{ij}=g^{im}g^{jn}F_{mn}\,.
  \end{equation}
 
Observe that now the components of two electromagnetic fields $F^{ij}$ and $F_{mn}$ are different functions of a point. In fact they can be treated as two independent physical fields. In such an approach, metric tensor comes from a relation between these independent fields, i.e., from a physical phenomen. 

Since $F^{ij}$ is antisymmetric, the inhomogeneous field equation
of (\ref{Max1}) yields a modified electric charge conservation law
\begin{equation}\label{Max02x}
\left(J^i\sqrt{-g}\right)_{,i}=0\,.
   \end{equation}
In fact, this equation is not a conservation law for the vector field $J^i$
itself. What is really conserved is the product of  $J^i$ with the root of the metric determinant. 
It means that a conserved electric current cannot be described by
a covariant vector field so a redefinition of this basic notion
is necessary. Instead of treating it as a vector field, the
electric current has to be considered as a weight $(+1)$
pseudo-vector density field 
\begin{equation}\label{Max02xx}
{\cal J}^i= J^i\sqrt{-g}\,. 
   \end{equation}
Also a
weight $(+1)$ pseudo-tensor density  of electromagnetic excitation
\begin{equation}\label{Max02xxx}
{\cal H}^{ij}=F^{ij}\sqrt{-g} 
 \end{equation}
 has to be involved. Under smooth
transformations of the coordinates $x^i\to x^{i'}$ with the
Jacobian $L={\rm det} (\partial x^{i'}/\partial x^{i})$, the
transformation law for these pseudo-tensorial quantities involves  an additional
factor $1/|L|$. In order to have a covariant field equation, this factor must be compensated.  The first order partial 
derivatives of the term $\sqrt{-g}$ makes the job and the whole equation is covariant. 

Consequently, the general covariant field equations  take the form
   \begin{equation}\label{Max1-2}
\epsilon^{ijkl}F_{jk,l}=0\,,\qquad {\cal H}^{ij}{}_{,j}={\cal J}^i\,,
  \end{equation}
  while the general covariant charge conservation law is written as
   \begin{equation}\label{Max1-3}
  {\cal J}^i{}_{,i}=0\,.
  \end{equation}
  The  constitutive relation between two basic fields takes the form 
  \begin{equation}\label{Max1-1x}
F^{ij}=\frac 12\chi^{ijmn}F_{mn}
   \end{equation}
   where the  constitutive pseudotensor
      \begin{equation}\label{Max1-3x}
\chi^{ijmn}=\sqrt{-g} \left(g^{im}g^{jn}-g^{in}g^{jm}\right)
\end{equation}
is involved. 

 Although the described modification serves the curvilinear
 coordinates on a flat manifold, it is well known to be enough
 also for description of the electromagnetic field in a curved spacetime of GR.
  Both field equations (\ref{Max1-2}) and the conservation
  law (\ref{Max1-3}) are general covariant even being written via
  the ordinary partial derivatives.


\vspace{0.5cm}

\noindent{\it C. Electrodynamics in anisotropic media. }
For an anisotropic media in a flat Minkowski space, Maxwell's
electrodynamics is described by two pairs of 3D vectors
$E_\a,B^\a$ and $D^\a,H_\a$, where the Greek indices are assumed
to obtain the spatial values, $\a,\b,\cdots=1,2,3$. In the 4D
notation, these vectors are assembled into two antisymmetric
tensors: the electromagnetic strength tensor  $F_{ij}$ with the
components
\begin{equation}\label{F-decomp-ten}
F_{0\a}=E_\a\,,\qquad F_{\a\b}=-\epsilon_{\a\b\g}B^\g\,,
\end{equation}
and  the electromagnetic excitation tensor $H^{ij}$ with the components
\begin{equation}\label{H-decomp-ten}
H^{0\a}=D^\a\,,\qquad H^{\a\b}=\epsilon^{\a\b\g}H_\g\,.
\end{equation}
In the 4D notation, the Maxwell field equations for the
electromagnetic field in an anisotropic media are written in the
form
\begin{equation}\label{Max-eq-anis}
\epsilon^{ijkl}F_{ij,k}=0\,, \qquad{ H}^{ij}{}_{,j}={ J}^i\,.
\end{equation}
An additional ingredient, the constitutive relation between two
electromagnetic tensors, $F_{ij}$ and $H^{ij}$,  describes the
characteristic properties of the media. For a wide range of
anisotropic   materials, a linear constitutive relation is a
sufficiently good approximation
\begin{equation}\label{const-rel-anis}
D^\a=\varepsilon^{\a\b}E_\b+\gamma^\a{}_\b B^\b\,,\qquad
H_\a=\mu^{-1}_{\a\b}B^\b+{\tilde{\gamma}}_\a{}^\b E_\b\,.
\end{equation}
The electromagnetic current conservation law $J^i{}_{,i}=0$ is a
consequence of the field equation (\ref{Max-eq-anis}). Notice also
that the equations (\ref{Max-eq-anis}) are invariant under
arbitrary constant linear transformations of the coordinates.
\subsection{ Premetric field equations}
The  models  accounted above show some similarity:
\begin{itemize}
\item  The electromagnetic field is described by two
2-nd order antisymmetric tensors. 
In differential form  notation, the field is represented by two 
2-nd order differential forms --- one twisted and one
untwisted.
\item All the models are described by similar systems of two first
order partial differential equations.
\item The antisymmetric tensorial fields are related  by a linear constitutive relation.
\item Even in vacuum electrodynamics, the metric of the manifold emerges only via a special 4-component tensor, i.e., it  plays only a secondary role.
\end{itemize}
The accounted similarity naturally leads to a premetric generalization of
the classical electrodynamics. For a comprehensive account of this
subject, see \cite{Birkbook}, \cite{HIO} and the references given
therein.

In the premetric  approach, two differential field equations for
two second order differential forms, the {\it electromagnetic
field strength} $F$ and the {\it electromagnetic excitation}
${\cal H}$ are postulated:
 \begin{equation}\label{preMax0}
dF=0\,,\qquad d{\cal H}={\cal J}\,.
  \end{equation}
In (\ref{preMax0}),  $F$ is an even (untwisted) differential
form. It does not change under arbitrary transformations of
coordinates. Alternatively,
 ${\cal H}$ and ${\cal J}$ are odd (twisted) differential forms.
 Under a change of coordinates with a Jacobian $L=det\left(L^i{}_j\right)$,
 they multiplied by the sign factor of $L$.
Namely such identification of the electromagnetic fields
guarantees the proper integral conservation laws for magnetic flux
and electric  current, see \cite{Birkbook}. Notice that the
domain of integration for a twisted form must carry a chosen
orientation.

Both  equations (\ref{preMax0}) are expressed via differential
forms thus they are manifestly invariant under arbitrary smooth
coordinate transformations.
 In a more general setting, see \cite{Birkbook}, these equations
 can be considered as consequences of two integral conservation laws:
 one for the magnetic flux and one for the electric current.

In a coordinate chart, we represent  the forms as
 \begin{equation}\label{preMax1}
F=\frac 12 \,F_{ij}dx^i\wedge dx^j \,,\qquad {\cal H}=\frac 12
\,{\cal H}^{ij}\epsilon_{ijmn}\,dx^m\wedge dx^n\,,
  \end{equation}
  while
    \begin{equation}\label{preMax1x}
{\cal J}= \frac 1{3!}\,{\cal J}^i\epsilon_{ijmn}\,dx^j\wedge
dx^m\wedge dx^n\,.
  \end{equation}
Thus, the components $F_{ij}$ constitute an ordinary antisymmetric
tensor while the components ${\cal H}^{ij}$ and ${\cal J}^{i}$ are
pseudotensor densities of weight $(+1)$.

Applying the exterior derivatives to (\ref{preMax1}) we rewrite 
the field equations (\ref{preMax0}) in the tensorial form
  \begin{equation}\label{preMax}
\epsilon^{ijkl}F_{jk,l}=0\,,\qquad {\cal H}^{ij}{}_{,j}={\cal J}^i\,.
  \end{equation}
Even being written via the ordinary partial
derivatives, these equations are covariant under arbitrary smooth
transformations of coordinates.
  Notice also that a metric tensor  or a connection are not involved
  in the construction above. One can even say that these structures
 are not  defined (yet) on the space.
 In this sense, the construction is premetric.  Particularly, in
 such approach, the covariant components of the field strength
tensor and the contravariant components of the excitation tensor cannot
 be introduced --- the indices cannot being raised or downed.
\subsection{Constitutive relation}
The system (\ref{preMax}) involves 8 equations for  12 independent
 variables so it is undetermined.
 Moreover,  the 2-form $H$ describes a  field generated
by a charged source, while the 2-form $F$ describes some other
 field which acts on a test charge. On a current  stage
of the construction, these two fields are formal and completely independent. 
It means that an interaction between two charges is not yet involved. 
In order to close the system and to involve an interaction, a constitutive relation between the
fields $F_{ij}$ and ${\cal H}^{ij}$ must be implicated. The simplest
choice of a local linear homogeneous relation
  \begin{equation}\label{ConstRel}
{\cal H}^{ij}=\frac 12 \chi^{ijkl}F_{kl}\,
\end{equation}
  is wide enough to describe the most observation data of the ordinary
  electrodynamics and  even involves some additional electromagnetic
  effects, such as axion, dilaton and skewon partners of photon, see\cite{Birkbook}.
  Also for the electromagnetism into the
  non-magnetized media,  the linear constitutive relation is a good
  approximation. For the nonlinear extensions of the premetric approach, see
  \cite{Obukhov:2002xa}. The non-local constitutive relations was considered recently  in \cite{Hehl:2007an}.

  Recall that the physical space is considered as a bare manifold
without metric or connection. All the  information on its
geometry is encoded into the constitutive pseudotensor
$\chi^{ijkl}$ which can depend of the time and position
coordinates.
  By definition, this pseudotensor
inherits the symmetries of the antisymmetric tensors $F_{ij}$,
${\cal H}^{ij}$. In particular,
\begin{equation}\label{ConstTenSym}
\chi^{ijkl}=\chi^{[ij]kl}=\chi^{ij[kl]}\,.
\end{equation}
Thus, in general, the  fourth order constitutive tensor
$\chi^{ijkl}$ has  36 independent components instead of $4^3$.
  Due to the Young diagrams analysis, under the group of
linear transformations, such a  tensor  irreducible decomposed 
 into three independent pieces
\begin{equation}\label{ConstTenReduction}
\chi^{ijkl}={}^{\tt (1)}\chi^{ijkl}+{}^{\tt (2)}\chi^{ijkl}+{}^{\tt
(3)}\chi^{ijkl}\,.
\end{equation}
The {\it axion part} (1 component) and the {\it skewon part}
(15 components) are
defined respectively as  
\begin{equation}\label{Axi+Skew}
{}^{\tt (3)}\chi^{ijkl}=\chi^{[ijkl]}\,,\quad
{}^{\tt (2)}\chi^{ijkl}=\frac
12\left(\chi^{ijkl}-\chi^{klij}\right)\,.
\end{equation}
The remainder ${}^{\tt (1)}\chi^{ijkl}$ is a {\it principal part}
of 20 independent components. One can also extract a scalar factor from the principle part which was identified  recently \cite{Hehl:2007an} with the dilaton partner  of electromagnetic field \cite{Bekenstein:1982eu},
  \cite{Bekenstein:2002wz}.

\subsection{Uniqueness and existence problem}

In this paper, we are studying the system of field equations
(\ref{preMax}) with a linear constitutive relation
(\ref{ConstRel}). The constitutive tensor   $\chi^{ijkl}$ is
irreducible decomposed to a sum of three independent pieces. In
absence of a metric tensor, a successive decomposition is not
available. Since the decomposition (\ref{ConstTenReduction}) is
irreducible, the difference pieces of the constitutive tensor can
appear alone or in certain linear combinations. Thus, in principle, we have several  non-trivial electromagnetic models. Not all of them,
however, confirm with the physical meaning of the field equations (\ref{preMax}).

In order to extract some physics meaningful models, we will apply the following requirements:

(i) {\it Hyperbolicity. } The free field must propagate in spacetime by waves. Thus
the partial differential equations has to be of the hyperbolic type.

(ii)  {\it Existence and uniqueness. } For physically meaningful sources, the  system 
(\ref{preMax}) with the  relation (\ref{ConstRel}) can serve as
a physical model only if it has a solution and this solution is unique.

It is easy to see that these requirements are not trivial. In
particular,  (ii) is not satisfied in the following case: 
Let only the axion part ${}^{\tt
(3)}\chi^{ijkl}$  be nonzero. Let it also be a constant. In this
case, the equations (\ref{preMax}) with the  relation (\ref{ConstRel}) are
inconsistent for ${\cal J}^i\ne 0$ and undetermined for ${\cal
J}^i=0$. Consequently an isotropic and homogeneous pure axionic
media is forbidden by the uniqueness and existence conditions.

One of the main  tasks of the current paper  is to establish under what
conditions a generic  constitutive tensor can represent a meaningful
physical system.

\section{Approximation and ansatz}
\subsection{Semi-covariant approximation}
  When the constitutive relation (\ref{ConstRel}) is substituted into
   the field equation (\ref{preMax}$_b$), we remain with
 \begin{equation}\label{sec-M1}
 \frac 12 \chi^{ijkl}F_{kl,j}+\frac 12
  \chi^{ijkl}{}_{,j}F_{kl}=
 {\cal J}^i\,.
 \end{equation}
The first term here describes how the electromagnetic field
changes  in a spacetime of  constant media characteristics.
Alternatively, the second term describes the temp of the
spacetime variation of the media characteristics for a constant
electromagnetic field. In this paper, we restrict to the {\it
geometrical approximation of optics}. In particular, we neglect
with the second term of (\ref{sec-M1}) relative  to the first
one. In other words, we restrict to a media which characteristics
change slowly on the characteristic distances of the change of the
fields. Note that such approximation is not always applicable. In particular, the    
Carroll-Field-Jackiw modification of electrodynamics \cite{Carroll:1989vb}, 
can be reformulated as a premetric electrodynamics \cite{Itin:2004za}, \cite{Itin:2007wz},\cite{Itin:2007cv} where the first term  of (\ref{sec-M1}) vanishes, while the birefringence effect comes namely from the second term.

 In the framework of the geometrical approximation, we remain with a system of 8 equations for 6 independent components of the electromagnetic field strength
$F_{ij}$
  \begin{equation}\label{MaxReduced}
\epsilon^{ijkl}F_{jk,l}=0\,,\qquad
\frac 12 \chi^{ijkl}F_{kl,j}={\cal J}^i\,.
  \end{equation}
In a special case of the Maxwell constitutive tensor,  the
approximation used here yields the inhomogeneous field equation
of the form
\begin{equation}\label{Max2Reduced}
g^{ik}g^{jl}F_{kl,j}= {\cal J}^i\,.
  \end{equation}
This is an approximation of the covariant equation (\ref{Max1})
when the derivatives of the metric tensor are considered to be
small relative to the derivatives of the electromagnetic  field.
Such a {\it semi-covariant approximation} is preserved for arbitrary
coordinate transformations with small spacetime derivatives.

  \subsection{Eikonal ansatz}
To describe the wave-type solutions of the field equations
(\ref{MaxReduced}), we consider  an {\it eikonal ansatz}.  Let the
electric current be given in the form
 \begin{equation}\label{eiko1}
J^i=j^ie^{\sigma}\,,
\end{equation}
and let the corresponded field strength be expressed as 
  \begin{equation}\label{eiko2}
F_{ij}=f_{ij}e^{\sigma}\,.
\end{equation}
Here the eikonal $\sigma$ is a scalar function of a spacetime
point. The tensors  $j^i$ and $f_{ij}$ are assumed to be slow
functions of a point. The derivatives  of $\sigma$ give the main
contributions to the field equations. Define the {\it wave
covector}
\begin{equation}\label{w-cov}
q_i=\frac{\partial \sigma}{\partial x^i}\,.
\end{equation}
In this approximation, the conservation law for the electric current $J^i{}_{,i}=0$ takes a form of an algebraic equation 
\begin{equation}\label{electr-cons}
j^iq_i=0\,.
\end{equation}
 Substituting (\ref{eiko1},
\ref{eiko2}) into the field equations (\ref{MaxReduced}) and
removing the derivatives of the amplitudes relative to the derivative of the eikonal function, we come to an algebraic system
  \begin{equation}\label{MainSystem}
 \epsilon^{ijkl}q_jf_{kl}=0\,, \qquad  \chi^{ijkl}q_jf_{kl}=2j^i\,.
 \end{equation}
 The same system was derived in \cite{Birkbook} by mean of
 Hadamar's discontinuity propagation method. This fact indicates that a simple 
 approximation used here is not less general than  the one used
 in \cite{Birkbook}.

Observe a remarkable property of (\ref{MainSystem}). If all the
quantities involved here are assumed to transform by  ordinary
tensorial transformation rules with arbitrary pointwise matrices,
both equations are preserved. In other word, these equations are
straightforward expanded to a general covariant system. This is
in spite of the fact that the approximations used in their
derivation are not covariant. This property is generic for
quasilinear systems which leading terms (the higher order derivatives
expressions) are linear and thus preserve their form even under
arbitrary pointwise transformations.

  \section{Characteristic equations}
  \subsection{The linear system}
  The approximation (\ref{MaxReduced}) and the wave-type ansatz
 (\ref{eiko1}, \ref{eiko2})
  yield a linear system (\ref{MainSystem}) of 8 equations for 6
independent variables.
  This algebraic system will serve as a starting point of our analysis.
  Observe first that the system  is not overdetermined.
  Indeed, when both equations  are multiplied
  by a covector $q_i$, they turn to trivial  identities,
  provided that  the electric current is conserved.
  Thus we have two linear relations between 8 linear
  equations (\ref{MainSystem}) for 6 independent variables.
  It means that the rank of the system (\ref{MainSystem}) is less or equal
  to six. The physical meaning of this system requires the rank to be exactly equal to 6. 
  Indeed, the unknowns $f_{kl}$ of this system are  physically
  measurable quantities. Thus, for an arbitrary conserved current $j^i$, they have to
  be determined  from (\ref{MainSystem}) uniquely.
  This  physical requirement puts a strong  algebraic
   constrain on the system (\ref{MainSystem}) --- {\it  it coefficients   must 
   form a matrix of a rank of 6.} In fact, it is a constrain on the
   components of the constitutive pseudotensor $\chi^{ijkl}$, which formal expression
   we will derive  in sequel.

\subsection{The homogeneous equation}

The homogeneous  equation (\ref{MainSystem}) is exactly the same
as in the standard Maxwell theory. We give
here a precise treatment of this equation mostly in order to establish the
notation and to illustrate the method used in  sequel.

  {\bf Proposition 1.} {\it A most general solution of a linear system
  \begin{equation}\label{LinMax1}
\epsilon^{ijkl}q_jf_{kl}=0
\end{equation}
  is expressed as
   \begin{equation}\label{sol1}
  f_{kl}=\frac 12\left(a_kq_l-a_lq_k\right)\,,
  \end{equation}
  where $a_k$ is an arbitrary covector.}
  \vspace{.3cm}

{\bf Proof:}
  The expression (\ref{sol1}) is evidently a solution  of (\ref{LinMax1}).
  In order to prove that it is a most general solution, we first
  notice that (\ref{LinMax1}) is a linear system of 4 equations for
  6 independent variables $f_{ij}$.
  The $4\times 6$ matrix of this system $a^{ikl}=\epsilon^{ijkl}q_j$
  with $ik=01,02,03,12,23, 31$ and $l=0,1,2,3$ is given by
   \begin{equation}\label{eq1-mat}
a^{ikl}=\left(\begin{array}{rrrrrr}
  0 &0&0&q_3&-q_2&q_1\\
  0&-q_3&q_2&0&0&-q_0\\
  q_3&0&-q_1&0&q_0&0\\
  q_2&q_1&0&-q_0&0&0
  \end{array}\right).
   \end{equation}
  The rows of this matrix satisfy a  linear relation
    \begin{equation}\label{eq1-mat2}
a^{ikl}q_i=0\,.
  \end{equation}
  Thus its rank is of 3 or less. If an arbitrary row is removed now from
  (\ref{eq1-mat}), the remaining three  columns are assembled in
  the echelon form.
  Thus the matrix (\ref{eq1-mat}) has exactly a rank of 3.
  Consequently, a general solution of (\ref{LinMax1}) has to involve
  $6-3=3$ independent parameters. This is exactly what is given in
   (\ref{sol1}).
Indeed, although the arbitrary covector $a_i$ has 4 independent
components, only 3 of them are involved in (\ref{sol1}). In
particular, the vector $a_i$ proportional to $q_i$ does not give
a contribution. Thus,  (\ref{sol1}) is a most general covariant
solution of (\ref{LinMax1}). $\blacksquare$

  \vspace{.3cm}
  It is well known that the homogeneous field equation (\ref{MaxReduced})
is solved in term of the  standard vector potential $A_i$.
  The covector $a_i$ appeared in (\ref{sol1}) is similar
  to the Fourier transform of  of the potential $A_i$. As ususal, this covector is arbirtrary if only 
 the homogeneous field equation is taken into account.

\subsection{The inhomogeneous equation}
Let us turn now to the inhomogeneous equation of (\ref{MainSystem}).
 Substituting the solution (\ref{sol1}),
 we arrive to   an algebraic system
  \begin{equation}\label{main2x}
  \chi^{ijkl}q_lq_ja_k=j^i\,.
  \end{equation}
  Observe that this is a system of 4 equations for 4 variables $a_k$.
  The matrix of this system
   \begin{equation}\label{main-matr}
  M^{ik}=\chi^{ijkl}q_lq_j\,,
  \end{equation}
will be refereed to as a {\it characteristic matrix}. We will see
in sequel that the wave propagation depends exactly on the
specific  combination of the components of the constitutive
pseudotensor $\chi^{ijkl}$ which are involved in $M^{ik}$.

  When the irreducible decomposition (\ref{ConstTenReduction})
  is substituted into the characteristic matrix $M^{ij}$, the completely antisymmetric axion part
  ${}^{\tt (3)}\chi^{ijkl}$ evidently does not contribute. 
  As for the other two pieces, the principle part is involved only in
  the symmetric part of the matrix $M^{ij}$, while the skewon part
  is involved only in its antisymmetric part. Formally, we can
  write
   \begin{equation}\label{M-decomp1}
  M^{(ik)}=M^{ik}\left(^{(\tt 1)}\chi\right)\,, \qquad
   M^{[ik]}=M^{ik}\left(^{(\tt 2)}\chi\right)\,.
  \end{equation}
So, the characteristic matrix $M^{ik}$ is irreducibly decomposed as
\begin{equation}\label{M-decomp2}
  M^{ik}=M^{(ik)}\left(^{(\tt 1)}\chi\right)+M^{[ik]}\left(^{(\tt
2)}\chi\right)\,.
  \end{equation}

 In the characteristic matrix notation,  equation (\ref{main2x}) takes the form
   \begin{equation}\label{main2-equation}
 M^{ik}a_k=j^i\,.
  \end{equation}
  The following two facts will play an important role in our
  analysis:

(1) {\it Gauge invariant condition.} Due to the antisymmetry of the constitutive pseudotensor  $\chi^{ijkl}$ in
its last two indices, an identity
   \begin{equation}\label{gauge}
  M^{ik}q_k=0\,
  \end{equation}
  holds true.
  It is a linear relation between the rows of the matrix $M^{ik}$. It
means that every solution of (\ref{main2x}) is defined only up to
an addition of a term $a_i\sim q_i$. This addition is evidently
unphysical since it does not contribute to the electromagnetic
strength. Consequently, relation (\ref{gauge}) has to be interpreted as a
 gauge invariant condition.

(2)  {\it Charge conservation condition.} Another evident identity for the matrix $ M^{ik}$ emerges
from the antisymmetry of the constitutive pseudotensor   $\chi^{ijkl}$ in its
first two indices
  \begin{equation}\label{charge}
  M^{ik}q_i=0\,.
  \end{equation}
  It is  a linear relations between the columns of the matrix $ M^{ik}$.
  Being compared with (\ref{main2-equation}), it yields  $j^iq_i=0$.
 Consequently, the relation (\ref{charge}) has to be
interpreted as a charge conservation condition.

  Thus we arrive to a some type of  a duality between the charge conservation
  and the gauge invariance. Note that  this duality is  expressed
  by a standard algebraic fact: For any matrix, the column rank and
  the row rank are equal one to another.

Due to the conditions indicated above, the rows (and the columns) of the matrix $M^{ij}$ are
linearly dependent, so its determinant is equal to zero. It can be
checked straightforwardly, but one has to apply here rather
tedious calculations.

\section{Dispersion relation}
\subsection{How it emerges}
  In the vacuum case of the  free electromagnetic waves, (\ref{main2x})
  takes a form of  a linear homogeneous system of four equations for
  four components of the covector $a_k$,
   \begin{equation}\label{main}
  \chi^{ijkl}q_lq_ja_k=0\,, \qquad {\rm or}\qquad M^{ik}a_k=0\,.
  \end{equation}
The gauge relation (\ref{gauge}) can be interpreted as a fact
that
  \begin{equation}\label{formsol}
  a_l=Cq_l
  \end{equation}
  is a formal solution of (\ref{main}).
  This solution does not contribute to the electromagnetic
field strength so it is unphysical. Hence, the formal system (\ref{main})  can describe  a
nontrivial wave propagation, only if it has an {\it additional  solution} which
must be linearly independent on (\ref{formsol}). Consequently (\ref{main}) must have at least two linear independent solutions. A known fact from the linear algebra that a  linear system
has two (or more) linearly independent solutions if and only if
the rank of the matrix $M^{ij}$ is of two (or less). In this case, the
adjoint matrix (constructed from the cofactors of $M^{ij}$) is equal to zero, $A_{ij}=0$.

 In order to present a formal expression of this fact we will start with a formula for the determinant of an arbitrary 4-th order matrix, 
   \begin{equation}\label{det}
det(M)=\frac 1{4!}
\epsilon_{ii_1i_2i_3}\epsilon_{jj_1j_2j_3}M^{ij}M^{i_1j_1}M^{i_2j_2}
M^{i_3j_3}\,.
   \end{equation}
   Here   the standard convention $\epsilon_{0123}=1$ is accepted.
The components of the adjoint matrix are expressed by the
derivatives of the determinant relative to the entries of the
matrix,
   \begin{equation}\label{Adj1}
A_{ij}=\frac{\partial\, det(M)}{\partial \,M^{ij}}\,.
\end{equation}
Explicitly,
  \begin{equation}\label{Adj1x}
A_{ij} =\frac 1{3!}
\epsilon_{ii_1i_2i_3}\epsilon_{jj_1j_2j_3}M^{i_1j_1}M^{i_2j_2}M^{i_3j_3}\,.
   \end{equation}
Consequently, we derived a physically motivated condition on the
  components of the constitutive pseudotensor $\chi_{ijkl}$.
  \vspace{.3cm}

  {\bf Theorem 2.} {\it The Maxwell system with a general linear
  constitutive relation has a non-trivial wave-type
  solution if and only if the adjoint of the matrix}
  $M^{ik}=\chi^{ijkl}q_lq_j$ {\it is equal to zero, i.e.,}
  \begin{equation}
  A_{ij} =0\,,
  \end{equation}
  or explicitly, 
 \begin{equation}\label{Adj1xx}
\epsilon_{ii_1i_2i_3}\epsilon_{jj_1j_2j_3}M^{i_1j_1}M^{i_2j_2}M^{i_3j_3}=0\,.
   \end{equation}
\vspace{.3cm}

  Since the adjoint matrix has 16 independent components, it seems that
we have to require, in general, 16 independent conditions. In
fact, the situation is rather simpler.
  The following  algebraic fact is
important in our analysis, so we present here its formal proof.
\vspace{.3cm}

  {\bf Proposition 3.} {\it Let 
  a square $n\times n$ matrix $M^{ij}$
  satisfies the relations
   \begin{equation}\label{matr-eq}
M^{ij}q_i=0\,,\qquad M^{ij}q_j=0\,,
  \end{equation}
  for an arbitrary nonzero $n$-covector $q_i$. 
  The adjoint  matrix $A_{ij}=Adj(M^{ij})$ is proportional to the tensor
  square of $q_i$, i.e.,
   \begin{equation}\label{Adj2}
A_{ij}=\lambda(q)q_iq_j\,.
   \end{equation}}
{\bf Proof:} Due to (\ref{matr-eq}), the rows (and the columns) of
$M^{ij}$ are linear dependent, so the matrix is singular and its
rank is equal to $n-1$ or less. 
 If the rank  is less than $(n-1)$, the adjoint matrix is
identically zero and (\ref{Adj2}) is satisfied trivially for
$\lambda=0$.

 Let the rank of $M^{ij}$ be equal to $(n-1)$. In this case,  $q_i$ is a {\it unique covector}
(up to a multiplication on a constant), that satisfies
(\ref{matr-eq}). It is well known that for a matrix of a rank of $(n-1)$, the  adjoint
$A_{ij}$ is a matrix of a rank of one.
  Moreover, an arbitrary  rank one matrix can be
  written as a tensor product of two covectors
   \begin{equation}\label{Adj1-gen}
A_{ij}=u_iv_j\,.
   \end{equation}
Let us show now that both these covectors must be proportional to
$q_i$. Indeed, the product of an arbitrary matrix with its adjoint is
equal to the determinant of the matrix times the unit matrix (this is the
generalized Laplace expansion theorem). In our case, $M^{ij}$ is
a singular matrix so
 \begin{equation}\label{Adj1-x-gen}
A_{ij}M^{ik}=A_{ij}M^{kj}=0\,.
 \end{equation}
 Substituting here (\ref{Adj1-gen}),  we have the  relations
   \begin{equation}\label{Adj1xx-gen}
u_iv_jM^{ik}=0\,\qquad u_iv_jM^{jk}=0\,,
   \end{equation}
  or, equivalently,
   \begin{equation}\label{Adj1xxx-gen}
u_iM^{ik}=0\,\qquad v_jM^{jk}=0\,.
   \end{equation}
  Comparing this pair of relations to (\ref{Adj2}) and remembering
that $q_i$ is unique up to a multiplication on a constant,  we
conclude that both covectors $u_i$ and $v_i$ are proportional to
$q_i$.
  Thus (\ref{Adj1}) indeed takes the form (\ref{Adj2}). $\blacksquare$
%

Consequently, in order to have a physically non-trivial
  vacuum wave-type solution, the system (\ref{main}) has to satisfy
  a unique scalar condition 
   \begin{equation}\label{disp1}
\lambda(q)=0\,.
   \end{equation}
   In fact, this is an expression for the principle {\it dispersion relation}. 
\subsection{Some basic properties of the dispersion relation}
Even without  an explicit expression for the function
$\lambda(q)$, we are ready now to derive some  characteristic
properties of the dispersion relation (\ref{disp1}).
 Some of these properties was recently derived in \cite{Birkbook} by involved straighforward calculations. 
In our approach,  these properties are immediate consequences
of the definition of the $\lambda$-function (\ref{Adj2}).

 \vspace{.3cm}

{\bf Corollary 4.} {\it { $\lambda(q)$ is a homogeneous 4-th order
polynomial of the wave covector $q_i$, i.e.,
  \begin{equation}\label{Conc1}
\lambda(q) ={\cal G}^{ijkl}q_iq_jq_kq_l\,,
   \end{equation}
   where ${\cal G}^{ijkl}$ is a pseudotensor independent on $q_i$.
 }}

Indeed, the adjoint matrix is a homogeneous polynomial of the six
order in $q_i$. By (\ref{Adj2}), after extracting the product $q_i q_j$ we remain
with a sum of terms every one of which is a  product of four  components of the
covector $q_i$. Since $\lambda(q)$ is a
pseudoscalar, ${\cal G}^{ijkl}$ is a pseudotensor.

 \vspace{.3cm}

 {\bf Corollary 5.} {\it {$\lambda(q)$ is a homogeneous 3-rd order
  polynomial of the constitutive pseudotensor  $\chi^{ijkl}$,
  i.e. it is of the form
 \begin{equation}\label{Conc2}
{\cal G}^{ijkl}=[\epsilon\epsilon\cdot(\chi\chi\chi)]^{ijkl}\,,
   \end{equation}
 where the dot symbol denotes the tensor contraction with 
 the Levi-Civita pseudotensor $\epsilon$.
 }}

Indeed, the adjoint matrix is a sum of terms which are cubic in
the matrix $M^{ij}$. Every such a term is a product of three
$\chi$'s which remains also after extracting the  product $q_i
q_j$ in the right hand side of (\ref{Adj2}).

 \vspace{.3cm}

 {\bf Corollary 6.} {\it { The equation $\lambda(q)=0$ defines a two-folded
  algebraic cone.} }

For a given set of vectors $K=\{x_1, x_2,\cdots\}$, an {\it algebraic
cone} is defined as a set of vectors
 $CK=\{Cx_1, Cx_2, \cdots\}$, where $C>0$ is an arbitrary number.
Due to the homogeneity of the dispersion relation (\ref{disp1}),
every its solution is defined up to a product on a constant. Thus
we have two folds of the cone  --- one for $C>0$ and the second
for $C<0$. This two-folded algebraic cone is a prototype of a light cone 
emergered by an Lorentz metric of vacuum electrodynamics or 
by optical metric of electromagnetism in a dielectric media. 
 \vspace{.3cm}

{\bf Corollary 7.} {\it{The axion part of the constitutive tensor
does not contribute to the function $\lambda(q)$.
  In other words,
   \begin{equation}\label{ax}
\lambda\left(^{\tt (1)}\chi+^{\tt (2)}\chi+^{\tt (3)}\chi\right)=
\lambda\left(^{\tt (1)}\chi+^{\tt (2)}\chi\right)\,.
   \end{equation}}}

Indeed, the axion part does not contribute to the  matrix
$M^{ij}$, so it does not appear in its adjoint.

 \vspace{.3cm}

{\bf Corollary 8.} {\it{The skewon part alone does not emerges  a
non-trivial dispersion relation.  In other words,  a relation}}
   \begin{equation}\label{skewon}
\lambda\left(^{\tt (2)}\chi\right)= 0\,
   \end{equation}
   {\it holds identically.  }

Indeed, in order to have a non-trivial (non-zero) expression for
$\lambda(q)$, the rank of the matrix $M^{ij}$ has to be equal to
three. The skewon par alone generates an antisymmetric matrix $M^{[ij]}$.
Since the rank of an arbitrary antisymmetric matrix is  even, the
skewon part alone does not emerges  a non-trivial dispersion
relation.

 \vspace{.3cm}

{\bf Corollary 9.} {\it{A non-trivial  (non-zero) dispersion
relation emerges only if the principle part of the constitutive
tensor is non-zero
   \begin{equation}\label{prin}
^{(1)}\chi\ne 0\,.
   \end{equation}}}
This is an immidiate result of the previous statements. 

\section{Dispersion relation in an explicit form}
\subsection{Covariant dispersion relation I}
   Our task now is to derive an explicit expression for  the dispersion relation. 
Recall that it is represented by a scalar equation 
  \begin{equation}\label{disp}
\lambda(q)=0\,,
   \end{equation}
where the function $\lambda(q)$ satisfies the equation 
   \begin{equation}\label{disp-1}
A_{ij}=\lambda(q)q_iq_j\,
   \end{equation}
   for the adjoint matrix $A_{ij}$ of the characteristic  matrix $M^{ij}$. 
To have an explicit expression for $\lambda(q)$, it is necessary  "to divide"
both sides of (\ref{disp-1}) by the product $q_iq_j$. Certainly
such a "division" must be produced in a covariant way. We will look first for a solution of this problem in  a special
coordinate system.  Let  a zeroth (time) axis be directed as a wave
covector, i.e.,  $q_0=q, q_1=q_2=q_3=0$. 
Substituting into (\ref{disp-1}) we have
\begin{equation}\label{disp-2}
\lambda(q) q^2=\frac 1{3!} \epsilon_{0i_1i_2i_3}
\epsilon_{0j_1j_2j_3}M^{i_1j_1}M^{i_2j_2}M^{i_3j_3}\,.
   \end{equation}
Due to the symmetry properties of the Levi-Civita pseudotensor,
 all  four-dimensional indices can be replaced
by the three-dimensional ones ($\a,\b=1,2,3$). So we get
\begin{equation}\label{Adj-new1}
\lambda(q) q^2=\frac 1{3!} \epsilon_{0\a_1\a_2\a_3}
\epsilon_{0\b_1\b_2\b_3}M^{\a_1\b_1}M^{\a_2\b_2}M^{\a_3\b_3}\,.
   \end{equation}
In the chosen system, the non-zero components of the matrix $M^{ij}$ are
  \begin{equation}\label{3D-M}
M^{\a\b}=\chi^{\a m\b n}q_mq_n=\chi^{\a 0\b 0}q^2\,.
  \end{equation}
Consequently,
\begin{equation}\label{Adj-new1n}
\lambda(q)=\frac 1{3!} \epsilon_{\a_1\a_2\a_3}
\epsilon_{\b_1\b_2\b_3}\chi^{\a_1 0\b_1 0}\chi^{\a_2 0\b_2
0}\chi^{\a_3 0\b_3 0}q^4\,.
   \end{equation}
 where the three dimensional Levi-Civita pseudotensor
 $\epsilon_{\a_1\a_2\a_3}=\epsilon_{0\a_1\a_2\a_3}$ is
  involved.
  The non-covariant dispersion relation takes the form \cite{Birkbook}
\begin{equation}\label{non-cov-disp}
\epsilon_{\a_1\a_2\a_3} \epsilon_{\b_1\b_2\b_3}\chi^{\a_1 0\b_1 0}\chi^{\a_2
0\b_2 0}\chi^{\a_3 0\b_3 0}q^4=0\,.
 \end{equation}
Note that this expression is only formal. Indeed, it is
equivalent to $q_0=q=0$, i.e, {\it all} the components of the wave covector are equal to zero. 
 This statement cannot represent any physical
meaningful property of the wave propagation. It is due to the
fact that the system with a zero wave covector cannot be realized by any
matter particles, i.e., there is not an observer with a 4-velocity equal to the wave covector.

  In \cite{Obukhov:2000nw}, \cite{Birkbook}, the equation
  (\ref{non-cov-disp})
was derived by  consideration of the three-dimensional determinant
of the system. In spite of the problems with it's physical neaning, the expression (\ref{non-cov-disp}) was used as a guiding form for a proper dispersion relation. In particular,    by physical intuition, this equation  was generalized to a covariant four-dimensional dispersion relation
\begin{equation}\label{cov-dis2}
\frac 1{4!}\epsilon_{ii_1i_2i_3}\epsilon_{jj_1j_2j_3}
\chi^{ii_1ja}\chi^{bi_2j_1c}
 \chi^{di_3j_2j_3}q_aq_bq_cq_d=0\,.
   \end{equation}
The $\lambda$-function can be read off from this equation as
   \begin{equation}\label{rub2}
\lambda(q)=\frac 1{4!}\epsilon_{ii_1i_2i_3}\epsilon_{jj_1j_2j_3}
\chi^{ii_1ja}\chi^{bi_2j_1c} \chi^{di_3j_2j_3}q_aq_bq_cq_d\,.
   \end{equation}
 Although, from a formal point of view, the described way of the  derivation of the covariant dispersion relation  is very much doubtful,
  the final result (\ref{cov-dis2}) turns to be correct, as we will show in sequel.
\subsection{Covariant dispersion relation II}
We will give now a pure covariant derivation of the scalar
function $\lambda(q)$ and of the corresponded dispersion relation. Differentiation of (\ref{Adj2}) relative
to the components of the covector $q_m$ yields
    \begin{equation}\label{calc1}
\frac {\partial A_{ij}}{\partial q_m}=\frac {\partial\lambda(q)
}{\partial q_m}q_iq_j+
\lambda(q)\left(\delta^m_iq_j+\delta^m_jq_i\right)\,.
   \end{equation}
Let us contract this equation  over the indices $m$ and $i$ and use the  Euler formula
for a fourth order homogeneous function $\lambda(q)$. Consequently,  we derive
    \begin{equation}\label{calc2}
\frac {\partial A_{ij}}{\partial q_i}=9\lambda(q)q_j\,.
   \end{equation}
  A second order derivative of this expression is given by
    \begin{equation}\label{calc3}
\frac {\partial^2 A_{ij}}{\partial q_i\partial q_m}=9\left(\frac
{\partial\,\lambda(q) }{\partial q_m}q_j+\lambda(q)\delta^m_j\right)\,.
   \end{equation}
 Summing now over the indices $m$ and $j$ and using once more
 the Euler formula, we derive
    \begin{equation}\label{calc4}
\lambda(q)=\frac 1{72}\,\frac {\partial^2 A_{ij}}{\partial q_i\partial q_j}\,.
   \end{equation}
Consequently, we have proved the following

\vspace{.3cm}

 {\bf Theorem 10.} {\it
For the Maxwell system with a general local linear
  constitutive relation,
the dispersion relation is given by}
     \begin{equation}\label{calc5}
 \frac {\partial^2 A_{ij}}{\partial q_i\partial q_j}=0\,.
   \end{equation}

\vspace{.3cm}

 In order to have an expression of the $\lambda$-function
 in terms of the matrix $M^{ij}$, we calculate the derivatives
 of the adjoint matrix
  \begin{eqnarray}\label{calc5a}
  \frac {\partial A_{ij}}{\partial
q_i}=\frac 1{2}\epsilon_{ii_1i_2i_3}\epsilon_{jj_1j_2j_3}
  \frac {\partial M^{i_1j_1}}{\partial q_i}M^{i_2j_2}M^{i_3j_3}\,.
   \end{eqnarray}
Substituting into (\ref{calc5}) and calculating the second order derivative, we get
  \begin{eqnarray}\label{calc6}
 \fl \lambda(q)=\frac 1{144} \epsilon_{ii_1i_2i_3}\epsilon_{jj_1j_2j_3}\left(\frac {\partial^2
M^{i_1j_1}}{\partial q_i\partial q_j}M^{i_2j_2}+2\frac {\partial
M^{i_1j_1}}{\partial q_i}\frac {\partial M^{i_2j_2}}{\partial
q_j}\right)M^{i_3j_3}\,.
\end {eqnarray}
This expression may be useful for actual calculations of the dispersion
relation for different media.

An explicit expression of the $\lambda$-function
via the constitutive pseudotensor is  calculated by the derivatives of
the matrix
 \begin{equation}\label{calc7-0}
 M^{ij}=\chi^{iajb}q_aq_b=-\chi^{iabj}q_aq_b=-\chi^{i(ab)j}q_aq_b\,.
 \end{equation}
 The first order derivative is given by
  \begin{equation}\label{calc7}
\frac{\partial M^{i_1j_1}}{\partial q_i}=\frac{\partial }{\partial
q_i}\left(\chi^{i_1mj_1n}q_mq_n\right)
  =-2\chi^{i_1(im)j_1}q_m\,.
\end{equation}
Hence, the second order derivative reads
  \begin{equation}\label{calc7x}
\frac{\partial^2 M^{i_1j_1}}{\partial q_i\partial q_j}=
-2\chi^{i_1(ij)j_1}\,.
    \end{equation}
  Consequently, the left hand side of (\ref{calc6}) takes the form
  \begin{eqnarray}\label{calc9}
\fl \lambda(q)=\frac 1{3\cdot 4!} \epsilon_{ii_1i_2i_3}\epsilon_{jj_1j_2j_3}
  \Big(-\chi^{i_1(ij)j_1}M^{i_2j_2}+
  4\chi^{i_1(ia)j_1}\chi^{i_2(jb)j_2}q_aq_b\Big)M^{i_3j_3}\nonumber\\
\fl\quad\,\,\,\,\,\,= \frac 1{3\cdot 4!} \epsilon_{ii_1i_2i_3}\epsilon_{jj_1j_2j_3}
  \Big(\chi^{i_1(ij)j_1}\chi^{i_2(ab)j_2}+
  4\chi^{i_1(ia)j_1}\chi^{i_2(jb)j_2}\Big)M^{i_3j_3}q_aq_b\nonumber\\
\fl\quad\,\,\,\,\,\,= \frac 1{3\cdot 4!} \epsilon_{ii_1i_2i_3}\epsilon_{jj_1j_2j_3}
  \Big(\chi^{i_1(ij)j_1}\chi^{i_2(ab)j_2}+
  4\chi^{i_1(ia)j_1}\chi^{i_2(jb)j_2}\Big)\chi^{i_3(cd)j_3}q_aq_bq_cq_d
  \,.
  \nonumber\\
\end {eqnarray}
  We finally have the covariant dispersion relation in the form
   \begin{equation}\label{calc10}
\fl\epsilon_{ii_1i_2i_3}\epsilon_{jj_1j_2j_3}
  \Big(\chi^{i_1(ij)j_1}\chi^{i_2abj_2}+
  4\chi^{i_1(ia)j_1}\chi^{i_2(jb)j_2}\Big)
  \chi^{i_3cdj_3}q_aq_bq_cq_d=0\,.
    \end{equation}
This equation turns out to be  equivalent to (\ref{cov-dis2}).
 The direct proof of this  fact was provided by Yu. Obukhov
 by a sequence of rather involved algebraic manipulations \cite{Ob}.

 The following decomposition represents the contribution of the skewon part into the dispersion relation.
Different forms of it can be found in \cite{Birkbook}.
 \vspace{.3cm}

 {\bf Proposition 12.} {\it
Due to the irreducible decomposition of the constitutive pseudo-tensor,
the dispersion relation $\lambda(\chi)=0$ is given by}
     \begin{eqnarray}\label{calc5x}
&&\fl\lambda\left({}^{(1)}\chi\right)+\frac 12\epsilon_{ii_1i_2i_3}\epsilon_{jj_1j_2j_3}
\frac {\partial^2 }{\partial q_i\partial q_j}\left[M^{(i_1j_1)}\left({}^{(1)}\chi\right)M^{[i_2j_2]}\left({}^{(2)}\chi\right)M^{[i_3j_3]}\left({}^{(2)}\chi\right)\right]=0\nonumber\\
   \end{eqnarray}

\vspace{.3cm}

 {\bf Proof:} The relation is follows straightforwardly when the decomposition (\ref{M-decomp2}) is substituted into
 (\ref{calc5}) and the antisymmetric terms are removed.   $\blacksquare$

\section{Maxwell electrodynamics  reinstated}
\subsection{Maxwell constitutive pseudotensor}
 When the premetric scheme is applied on a manifold with a
prescribed metric tensor $g_{ij}$, the standard Maxwell
electrodynamics  (\ref{Max1}) is
  reinstated by substitutions
    \begin{equation}\label{Max-substitution}
  {\cal H}^{ij}=\sqrt{-g}g^{im}g^{jn}F_{mn}=\frac 12\, \sqrt{-g}
\left(g^{im}g^{jn}-g^{in}g^{jm}\right)F_{mn}
  \end{equation}
Recall that the electric current vector density is expressed as
\begin{equation}\label{J-substitution}
{\cal J}^i=J^i\sqrt{-g}\,.
  \end{equation}
The constitutive relation (\ref{Max-substitution}) corresponds to
a choice of  a special {\it Maxwell-Lorentz  constitutive pseudotensor}
  \begin{equation}\label{M-con}
  ^{\tt{(Max)}}\chi^{ijkl}= \sqrt{-g}
  \left(g^{ik}g^{jl}-g^{il}g^{jk}\right)\,.
  \end{equation}
 We use here a system of units in which a constant with the dimension
 of an admittance (denoted by $\lambda_0$ in \cite{Birkbook})
 is taken to be equal to one.
When ({\ref{M-con}) is substituted in (\ref{preMax}) we return to
the standard Maxwell electrodynamics system  (\ref{Max1}).

\subsection{Dispersion relation}
Let us derive the standard  metrical dispersion relation. 
The characteristic matrix corresponded to the constitutive pseudotensor
  (\ref{M-con}) takes the form
    \begin{equation}\label{Max-matr}
  M^{ik}=\sqrt{|g|}\left(g^{ik}q^2-q^iq^j\right)  \,,
  \end{equation}
  where the notations $q^2=g^{ij}q_iq_j$ and $q^i=g^{im}q_m$ are used.
  The adjoint of this matrix is calculated straightforwardly
  \begin{eqnarray}
 \fl A_{ij} =\frac 1{3!}
\epsilon_{ii_1i_2i_3}\epsilon_{jj_1j_2j_3}M^{i_1j_1}M^{i_2j_2}M^{i_3j_3}\nonumber\\
\fl=\frac 1{3!}\sqrt{|g|^3}\epsilon_{ii_1i_2i_3}\epsilon_{jj_1j_2j_3}\left(g^{i_1j_1}q^2-q^{i_1}q^{j_1}\right)
\left(g^{i_2j_2}q^2-q^{i_2}q^{j_2}\right)\left(g^{i_3j_3}q^2-q^{i_3}q^{j_3}\right)\nonumber\\
\fl=\frac 1{3!}\sqrt{|g|^3}\epsilon_{ii_1i_2i_3}\epsilon_{jj_1j_2j_3}\left(g^{i_1j_1}g^{i_2j_2}g^{i_3j_3}q^6-
3g^{i_2j_2}g^{i_3j_3}q^{i_1}q^{j_1}q^4\right)\nonumber\\
\fl=\frac 1{3!}\sqrt{|g|^3}\left[-6g_{ij}q^6+6(g_{ij}g_{i_1j_1}
-g_{ij_1}g_{i_1j})q^{i_1}q^{j_1}q^4\right] \,.
  \end{eqnarray}
  The first two terms are cancel one another so  the adjoint  matrix remains in the form
  \begin{equation}\label{Max-adj}
A_{ij} =-\sqrt{|g|^3}q_iq_jq^4\,.
\end{equation}
Correspondingly, 
 \begin{equation}\label{Max-lambda}
\lambda =-\sqrt{|g|^3}q^4\,
\end{equation}
and the dispersion relation takes the  standard form
\begin{equation}\label{Max-disp}
q^2=0\qquad \Longleftrightarrow \qquad g^{ij}q_iq_j=0\,.
\end{equation}
From (\ref{Max-lambda}) we also deduce that the pseudotensor appeared in (\ref{Conc1},\ref{Conc2}) 
takes the form
\begin{equation}\label{Max-G}
{\cal G}^{ijkl}= -\frac 12 \sqrt{|g|^3}\left(g^{ij}g^{kl}+g^{ik}g^{jl}\right)\,.
\end{equation}

\section{Results and discussion}
Premetric electrodynamics can be viewed as a general framework for description of a wide class of electromagnetic effects. In this paper we discussed the geometric optics approximation of the wave propagation in this model. Due to the standard procedure of partial differential equations theory, such approximation represents the leading contribution to the corresponded solutions. We derived a covariant dipersion relation and showed that this relation represents the existance of the wave-type solution in the premetric electromagnetic model. It should be noted that our expression of the  covariant dispersion relation is not less complicated that the one represented in literature, \cite{Birkbook}. An advantage of  our approach is that we give an straightforward  covariant procedure how the dispersion relation can be derived for various constitutive relations. For this, one does not need to deal with the explicit covariant formula at all. 
It is enough to construct the characteristic matrix $M^{ij}$ and calculate it's adjoint $A_{ij}$.  Due to the proposition, proved in the paper, the extra factors $q_iq_j$ are separated from $A_{ij}$ and the remain part is the essential term of the dispersion relation. We have shown, how this procedure works in the case of the standard Maxwell 
constitutive relation. 
The problem of uniqueness of the wave-type solution in the premetric electromagnetic model is related to another principle notion  --- the photon propagator, \cite{Itin:2007av}, \cite{Itin:2007cv}.  A detailed consideration of this quantity and of it's relation to the uniqueness problem will be represented in a separate publication.


\section*{Acknowledgment}
  I would like to thank Friedrich Hehl and Yuri Obukhov  for most fruitful
 discussions.

\end{document}